
\magnification=\magstep1
\baselineskip=2\normalbaselineskip
\parskip 10 pt
\centerline{APPROXIMATE SELF-CONSISTENT MODELS FOR TIDALLY TRUNCATED}
\centerline{STAR CLUSTERS}
\bigskip

\centerline{D.C. Heggie \& N. Ramamani\footnote{*}{Now at Royal
Greenwich Observatory, Madingley Road, Cambridge CB3 0EZ.\break
RAMA@srf.rgo.ac.uk}}

\medskip
\centerline{University of Edinburgh,}
\centerline{Department of Mathematics and Statistics,}
\centerline{King's Buildings,}
\centerline{Edinburgh EH9 3JZ}
\centerline{d.c.heggie@ed.ac.uk}
\vfill
\eject

\centerline{\bf Abstract}
This paper generalises King's models for
tidally truncated star clusters by including approximately
the non-spherical symmetry of the tidal field and the
resulting non-spherical distortion of the cluster.

\vfill
\eject

\noindent{\bf1.  Introduction }

The problem of constructing a model of a stellar
system such as a globular cluster has a long history. In
the last twenty years or so, however, much the most widely
used models, at least in this context, are King's models
(King 1966) and various extensions of them.

Several factors have led to the popularity of
these models. First, they are very easy to compute.
Second, they are successful. Even the single mass models
are an excellent fit to the light distribution of most
observed galactic globular clusters (e.g. Illingworth
\& Illingworth 1976). Third, they are easily extended
(with suitable assumptions) to include a spectrum of
masses or anisotropy in the distribution of velocities,
and indeed such models are the standard means of interpreting
radial velocity data (e.g. Gunn \& Griffin 1979, Meylan
1988). Fourth, they approximately incorporate essential
aspects of theoretical stellar dynamics: (i) Jeans'
theorem (ii) relaxation effects and (iii) tidal truncation.

It is the aim of this paper to remove some
approximations involved in these models, especially with
regard to (i) and (iii).
In the construction of King's models the (specific)
energy of a star is taken as $v^2/2+\Phi_1$ where $v$ is the
speed and $\Phi_1$ is the potential due to the cluster. In
fact, each star is moving in the potential field of both
the cluster and the galaxy. Even so, the energy is not
conserved, because $v$ is measured in an accelerating frame
and centrifugal terms must be added.

In this paper, then, we shall construct a series of models analogous to King's
series, but approximately making allowance for the non-spherical symmetry of
the
tidal field and the non-conservation of energy. The method we use is an
expansion of the potential in spherical harmonics, though truncated at very low
order. We also use the same `lowered' Maxwellian distribution function as in
King's models, except that it is now a function of the Jacobi integral, which
is
a conserved quantity when the potential is in uniform rotation.  Thus, we
envisage a cluster whose barycentre rotates on a circular orbit in the plane of
symmetry of an axisymmetric galaxy. A shortened version of our results is
presented in Heggie \& Ramamani 1993, and we would like to mention that a
similar set of models is discussed by Weinberg (1993).

The outline of this paper is as follows. In section
2, we present the necessary preliminaries. Section 3 describes
the numerical method. In section 4 results are discussed for
two cases (i) when the galaxy is modelled as a point mass
and (ii) when the tidal field is modelled using observational
data from the vicinity of the sun.

\noindent{\bf2.  Theory of a tidally perturbed model star cluster }

\noindent2.1.  Case of a point mass perturber

The cluster under consideration has mass $M_1$, core radius $r_c$ and a
limiting radius of order $r_e$.  It contains stars of equal mass and
these stars will have an isotropic velocity distribution with respect to a
rotating frame of reference which will be described shortly.  The cluster moves
in a circular orbit of radius $D$ around the Galaxy, which is represented
by a point mass of mass $M_2$.  The changes in the Galactic potential
across the cluster are small for $D \gg r_e$, and the field of the
Galaxy is adequately represented in a linear approximation.

We choose a coordinate system rotating with an angular velocity equal to
that of the orbital motion of the Galaxy-cluster system, and choose its
origin to be at the centre of mass of this system.  The $x$-axis is chosen
to be pointing towards the Galaxy, the $y$-axis perpendicular to the
$x$-axis in the orbital plane in the sense of the orbital motion, and the
$z$-axis perpendicular to the orbital plane in the right handed sense.  We
denote all attributes of the cluster and the Galaxy with the suffices 1
and 2, respectively, and the central values of the cluster with the
suffix $c$.

A star in the cluster, at a point ${\bf r} = (x,y,z)$, has an integral
of motion $J$, the Jacobi constant, which is given by
$$
J = \Phi_1 + \Phi_2 - {1\over2}\omega^2(x^2 + y^2) +
{1\over2}v^2,\eqno(1)
$$
where $\Phi_1, \Phi_2$ are the potentials at ${\bf r}$ due to the cluster
and the Galaxy, respectively, $\omega$  is the angular velocity of the frame
of reference, and $v$ is the speed of the star in the rotating frame.
Writing the kinetic energy per unit mass as $T = {1\over2}v^2,$ we have
$J = U + T,$ where
$$
U = \Phi_1 + \Phi_2 - {1\over2}\omega^2(x^2 + y^2)\eqno(2)
$$
$$
\simeq \Phi_1 -{GM_2\over D}  -
{GM_2/D^3}(x_1^2 - {1\over 2}y_1^2 - {1\over 2}z_1^2)
- {GM_2\over 2D}
- {GM_2/2D^3}(x_1^2 + y_1^2)\eqno(3)
$$
where $x_1, y_1, z_1$ are the coordinates of the star with respect to
the centre of mass of the cluster, with the directions of the coordinate
axes chosen as before.  Expression (3) is obtained on the assumption
that $M_2 \gg M_1$, and $r_1 \ll D$.  The constant terms here are
irrelevant in what follows, by redefinition of the zero-point of the
potential.

{}From Jeans' theorem, the phase-space distribution function
(mass-density) for the stars of the cluster can be taken as any function
of $J$, and in order to make as close an analogy as possible with King's
models we take
$$
f({\bf r,v}) = K(\exp(-2j^2J) - \exp(-2j^2J_e)),\eqno(4)
$$
where $J_e$ is the value of $J$ at the limiting zero velocity surface,
at the edge of the cluster, and $K,~j$ are constants.
Then the density distribution is, as usual,
$$
\rho = \int_0^{v_e} f({\bf r,v}) 4\pi v^2dv, \eqno(5)
$$
where $v_e$    is the maximum speed at the point ${\bf r} = (x_1,y_1,z_1)$,
which is
given by
$$
v_e^2 = - 2(U - J_e).\eqno(6)
$$

We now exploit the arbitrariness in the zero-point of the potential by
supposing that $J_e = 0$.  Then it is easy to see that the density takes
the form
$$
\rho = \rho_c R(j^2U),\eqno(7)
$$
where $\rho_c$ is the central density and $R$ is expressible in terms of the
error function, exactly as in
the usual theory of King models (King 1966).  Where the theory differs
is in the expression for $U$, which may be written
$$
U = \Phi_1 + Q,\eqno(8)
$$
where $Q$ is the sum of the quadratic terms in expression (3).  Now if
these terms were neglected we would obtain King's models, and $\Phi_1$
would be the corresponding potential.  If the tidal field, represented
by $Q$, were weak, then $\Phi_1$ would differ slightly.  We therefore
write
$$
\Phi_1 = \Phi_k + \delta\Phi, \eqno(9)
$$
where $\Phi_k$ is the potential of the corresponding King model, and
$\delta\Phi$ is the change in the potential due to the distortion of the
cluster by the tidal field.  As usual, potential and density are related
by Poisson's equation:
$$
\nabla^2\Phi_1 = 4\pi G\rho(U).\eqno(10)
$$
Here is another difference between these models and King's:  $\rho$ depends
on $U$, not on $\Phi_1$.

Now we take the important step of linearising with respect to the deviation
from the King model, writing
$$
\rho(U) = \rho(\Phi_k) + (\delta\Phi + Q){d\rho\over d\Phi_k} +
\ldots.\eqno(11)
$$
Then from eqs. (9) - (11) it follows that
$$
\nabla^2\Phi_k = 4\pi G\rho(\Phi_k)\eqno(12)
$$
and
$$
\nabla^2\delta\Phi = 4\pi G {d\rho(\Phi_k)\over d\Phi_k}\{\delta\Phi + Q\}.
\eqno(13)
$$
Eq. (12) is the same as eq.(12) in King (1966).  We now write
$$
F(r) = 4\pi G{d\rho(\Phi_k)\over d\Phi_k},\eqno(14)
$$
$$
\delta\Phi(r,\theta,\phi) = \sum_0^\infty \delta\Phi_n(r)S_n(\theta,\phi)
\eqno(15)
$$
and
$$
Q = \sum Q_n(r)S_n(\theta,\phi),\eqno(16)
$$
where $r,~\theta,~\phi$ are spherical polar coordinates with origin at
the cluster centre, and the $S_n(\theta,\phi)$ are a complete set of surface
spherical
harmonics.  Eq.(13) can then  be written as
$$
[\nabla^2 - F(r)]\sum\delta\Phi_n(r)S_n(\theta,\phi) =
\sum F(r)Q_n(r)S_n(\theta,\phi),
$$
which implies
$$
\nabla^2\delta\Phi_n(r) - {n(n+1)\over r^2}\delta\Phi_n(r) -
F(r)\delta\Phi_n(r) = F(r)Q_n(r).\eqno(17)
$$

In fact only a few terms of these series are needed.   We have
$$
x_1 = r\sin\theta\cos\phi,~~
y_1 = r\sin\theta\sin\phi
$$
and
$$
z_1 = r\cos\theta,
$$
whence eqs.(3) and (8) give
$$
Q = -{GM_2\over D^3}r^2\{{1\over3}S_0(\theta,\phi) + S_2(\theta,\phi)\},
\eqno(18)
$$
where $S_0(\theta,\phi) \equiv 1$ and
$$
S_2(\theta,\phi) = -{5\over6}P_2^0(\cos\theta) + {1\over4}P_2^2(\cos\theta)
\cos2\phi.\eqno(19)
$$
Here $P_n^m$  are Legendre's associated polynomials of degree  $n$  and
order  $m$.

Since $Q$ has only zero- and second-order spherical harmonics, we get
only two relevant equations, with  $n = 0$  and $n = 2$, from eq.(17).
They are
$$
{d^2(\delta\Phi_0)\over dr^2} + {2\over r}{d(\delta\Phi_0)\over dr}
- 4\pi G {d\rho(\Phi_k)\over d\Phi_k}\delta\Phi_0 =
-4\pi G{d\rho(\Phi_k)\over d\Phi_k}.{GM_2\over D^3}{r^2\over3}\eqno(20)
$$
and
$$
{d^2(\delta\Phi_2)\over dr^2} + {2\over r}{d(\delta\Phi_2)\over dr}
-\{{6\over r^2} + 4\pi G {d\rho(\Phi_k)\over d\Phi_k}\}\delta\Phi_2 =
-4\pi G{d\rho(\Phi_k)\over d\Phi_k}.{GM_2\over D^3} r^2.\eqno(21)
$$
Since $S_0(\theta,\phi) = 1$ the quantity $\delta\Phi_0$ is the
spherically symmetric part of the change in the potential which arises
due to the distortion of the cluster, and $\delta\Phi_2$ multiplied by
$S_2$ gives the non-spherical part.  It is worth pausing here to note
the shape of this distortion.  Indeed from eq.(19) we have
$$
S_2(\theta,\phi) = {5\over12} + {3\over4}\sin^2\theta\cos 2\phi -
{5\over4}\cos^2\theta.
$$
Thus $S_2$ is largest along the $x_1$-axis and smallest along the $z_1$-axis.
The relevant values are $S_2(\pi/2,0) = 7/6$,
$S_2(\pi/2,\pi/2) = -1/3$ and $S_2(0,0) = -5/6.$

\noindent2.2 Jacobi integral for a galactic disk

Chandrasekhar (1942, p.217) gives the equations
$$
m(\ddot\xi - 2\omega_c\dot\eta + \alpha_1\xi) = -\partial\Omega/\partial\xi,
$$
$$
m(\ddot\eta + 2\omega_c\dot\xi) = -\partial\Omega/\partial\eta,
$$
$$
m(\ddot\zeta + \alpha_3\zeta) = -\partial\Omega/\partial\zeta,
$$
for the motion of a star of mass $m$ moving in the potential $\Omega$ of a
cluster.  The coordinates $(\xi,\eta,\zeta)$ have the same meaning as
$(x_1,y_1,z_1)$ in \S2.1, $\omega_c$ is $\omega$, $\alpha_1$ and
$\alpha_3$ are constants, and
subscripts $i$ have been dropped.  If $\Omega = \Omega(\xi,\eta,\zeta),$
i.e. we suppose the potential due to the cluster is time-independent in
the rotating frame, then multiplying these equations by $\dot\xi,
\dot\eta, \dot\zeta,$ respectively, and adding, gives the Jacobi
integral
$$
J = {1\over2}(\dot\xi^2 + \dot\eta^2 + \dot\zeta^2) +
{1\over2}\alpha_1\xi^2 + {1\over2}\alpha_3\zeta^2 + \Omega/m.\eqno(22)
$$
Denoting by $B$ the galactic potential, Chandrasekhar (p.217 again)
gives the definitions
$$
\alpha_1 = {\partial^2B\over\partial\varpi^2} - {1\over\varpi}{\partial
B\over \partial\varpi},\eqno(23)
$$
$$
\alpha_3 = {\partial^2B\over\partial z^2},\eqno(24)
$$
where $z$ is height above the galactic plane and $\varpi$ is the
distance from the symmetry axis of the galaxy.  Using Poisson's equation
in cylindrical polars (eq. 4.528 in Chandrasekhar) we find that
$$
\alpha_3 = 4\pi G\rho_G - {\partial^2B\over\partial\varpi^2} -
{1\over\varpi}{\partial B\over \partial\varpi},
$$
where $\rho_G$ is the galactic density.
{}From the first of eqs.(5.501) in Chandrasekhar we deduce that
$$
\alpha_3 = 4\pi G\rho_G - {2\Theta\over\varpi}{d\Theta\over d\varpi},
$$
where $\Theta$ is the local speed of circular galactic motion.
Hence,  by eq.(1.429) in Chandrasekhar,
$$
\alpha_3 = 4\pi G\rho_G - 2\omega(A + B),\eqno(25)
$$
where $A$ and $B$ are Oort's constants.
Similarly
$$
\alpha_1 = 4A(B - A),\eqno(26)
$$
by Chandrasekhar's eq.(5.615).

The Jacobi integral has exactly the same form as in the case of a point
mass $M_G$ at distance $D$.  In this case
$B = - GM_G/(\varpi^2 + z^2)^{1/2}$ and $\rho_G = 0$, whence $\alpha_1 =
-3GM_G/D^3$ (by eq.(23), where the potential derivatives are evaluated at
$z = 0, \varpi = D$) and $\alpha_3 = GM_G/D^3$.  Thus the Jacobi
integral is
$$
J = {1\over2}(\dot\xi^2 + \dot\eta^2 + \dot\zeta^2) -
{3\over2}{GM_G\over D^3}\xi^2 + {1\over2}{GM_G\over D^3}\zeta^2 + \Omega/m.
$$

This is essentially the same result as eq.(3) of section 2.1.

Returning to the general case (eq.(22)) and defining $Q$ by analogy with
eq.(8), we have
$$
Q = {1\over2}\alpha_1\xi^2 + {1\over2}\alpha_3\zeta^2
$$
where $\xi,\zeta$ have the same meaning as $x_1,z_1$ in \S2.1. Hence
$$
Q = {1\over3} \alpha_1r^2({1\over3}S_0+S_2)
$$
where now
$$
S_0={3\over2}(1+{\alpha_3\over\alpha_1})P_0^0\eqno(27)
$$
and
$$
S_2=({\alpha_3\over\alpha_1}-{1\over2})P_2^0+{1\over4}P_2^2\cos2\phi.
$$

\noindent {\bf3.  Numerical method}

In order to construct models it is necessary to solve eqs.(12), (20) and (21)
simultaneously throughout the cluster by numerical methods.  Following
King (1966), we choose scaled variables $W \equiv -2j^2\Phi_k$ and $R
\equiv r/r_c$, where $r_c$ is the core radius defined below.  Analogously the
perturbations in the potential are scaled
to give $y_3 = -2j^2\delta\Phi_0$ and $y_5 = -2j^2\delta\Phi_2.$
For the sake of definiteness we consider the case of \S2.1, and define
$$
\rho_1 = {M_2\over {4\over3}\pi D^3},
$$
which is a
measure of the density of the perturbing Galaxy. Following King
again, we take the core radius as satisfying the relation $8\pi G
j^2\rho_cr_c^2 = 9.$ With $R$ as the independent variable, the equations then
give
$$
{d^2W\over dR^2} + {2\over R}{dW\over dR} = - 9{\rho(W)\over \rho_c},\eqno(28)
$$
$$
{d^2y_3\over dR^2} + {2\over R}{dy_3\over dR} + 9{d(\rho/\rho_c)\over dW}y_3
= - 9R^2{\rho_1\over\rho_c}{d(\rho/\rho_c)\over dW},\eqno(29)
$$
and
$$
{d^2y_5\over dR^2} + {2\over R}{dy_5\over dR} - {6\over R^2}y_5
+ 9{d(\rho/\rho_c)\over dW}y_5
= - 27R^2{\rho_1\over\rho_c}{d(\rho/\rho_c)\over dW},\eqno(30)
$$
where, following King,
$$
{\rho\over\rho_c} = {\exp(W)\int_0^W\exp(-\eta)\eta^{3/2}d\eta\over
\exp(W_c)\int_0^{W_c}\exp(-\eta)\eta^{3/2}d\eta}\eqno(31)
$$
and
$$
(\rho/\rho_c)^\prime\equiv{d(\rho/\rho_c)\over dW} =
{\rho\over\rho_c} + {W^{3/2}\over
\exp(W_c)\int_0^{W_c}\exp(-\eta)\eta^{3/2}d\eta}.\eqno(32)
$$
Here, as usual, a subscript $c$ denotes a core or central value.

Since $\rho$ depends only on W, eqs.(28) to (30) may be conveniently
transformed with the choice of $W$ as the independent variable.  We also
use $\ln(1+R^2)$ rather than $R$ as the dependent variable in
transforming eq.(28), because this varies nearly linearly with $W$ both
near $R = 0$ and in an isothermal halo. Thus
the equations to be integrated take the form
$$\eqalignno{
{dy_1\over dW} &= y_2&(33)\cr
{dy_2\over dW} &= \left({y_2\over2R}\right)^2\left
(6 + 2R^2 + 9{\rho\over\rho_c}
(1 + R^2)^2y_2\right)&(34)\cr
{dy_3\over dW} &= y_4&(35)\cr
{dy_4\over dW}&= {9\over 4}{(1 + R^2)^2\over R^2}y_2^2
\left\{- R^2{\rho_1\over\rho_c}{d(\rho/\rho_c)\over dW} -
y_3{d(\rho/\rho_c)\over dW} + {\rho\over\rho_c}y_4\right\} &(36)\cr
{dy_5\over dW} &= y_6 {\rm\ and}&(37)\cr
{dy_6\over dW}&= {3\over 4}{(1 + R^2)^2\over R^2}y_2^2
\left\{{2y_5\over R^2}  - 9R^2{\rho_1\over\rho_c}{d(\rho/\rho_c)\over dW} -
3y_5{d(\rho/\rho_c)\over dW} + 3{\rho\over\rho_c}y_6\right\} &(38)\cr
}$$
where $y_1 = \ln(1+R^2)$ and the additional variables $y_2, y_4$ and $y_6$ are
essentially
defined by eqs.(33), (35) and (37).  Thus the original set of three
second-order equations has been transformed to a set of six equations of
first order.

The numerical treatment of these equations is straightforward.
The above system can be written as
$$
{d{\bf y}\over dx} = {\bf g}(x,{\bf y})
$$
and can be discretised as
$$
{{\bf y}_{i+1} - {\bf y}_i\over\Delta x} = {\bf g}(\bar x,\bar{\bf
y}),\eqno(39)
$$
where $\bar x = (x_i + x_{i+1})/2,$ etc., and a subscript $i$ indicates
the value at the $i$-th mesh point.

The boundary conditions at the centre of the system can be obtained
using the following developments.  Let
$$\eqalign{
W - W_c &= aR^2 + bR^4 + \ldots\cr
y_3 &= a_0R^2 + b_0R^4 + \ldots\cr
y_5 &= a_2R^2 + b_2R^4 + \ldots\cr
}\eqno(40)
$$
Substitution of eqs.(40) into eqs.(28 - 30) is found to give the results
$$\eqalign{
a &= -3/2,\cr
b &= {27\over40}(\rho/\rho_c)^\prime,\cr
a_0 &= 0,\cr
b_0 &= -{9\over20}{\rho_1\over\rho_c}(\rho/\rho_c)^\prime,{\rm\ and}\cr
b_2 &= -{9\over14}(\rho/\rho_c)^\prime(a_2 + 3{\rho_1\over\rho_c}),\cr
}$$
where $(\rho/\rho_c)^\prime$ is evaluated at the centre, and  $a_2$  is yet to
be determined. (It must be solved for along with the
solution of the differential equations.)  From these results it is easily found
that the boundary conditions at the centre are
$y_2 = -2/3, y_6 = -2a_2/3,$ and $y_1 = y_3 = y_4 = y_5 = 0.$
At the outside of the cluster $\delta\Phi_2$ should join smoothly to the
exterior (vacuum) solution proportional to $R^{-3}$, whence
$$
y_6 = -{3\over2}{1+R^2\over R^2} y_5y_2.
$$
Eqs.(39), along with the above boundary conditions, were solved
numerically using the Newton-Raphson method.

As a check of the numerical method, at least as far as the equations for
$y_1$ and $y_2$ are concerned, King's models were regenerated using our
code, and good agreement was obtained with the results of King (1966).
This yields a one-parameter family of solutions, depending on $W_c$.
But the system of eqs.(33)-(38) also possesses another dimensionless
parameter, viz.  $\rho_1/\rho_c$.  In physical terms the number of
parameters can be understood by considering the number of radii which
characterise each solution.  In fact there are three: the core radius,
$r_c$; the radius of the system, $r_e$; and, finally, the radius of the
last closed zero-velocity surface, on which the Lagrangian points of the
cluster-Galaxy potential lie.  These three radii give two independent
dimensionless ratios, i.e.  two parameters.

A solution of the equations having been obtained for given values of the
two parameters $W_c$ and $\rho_1/\rho_c$, the functions $\Phi_k, \delta\Phi_0$
and
$\delta\Phi_2$ can be determined from the runs of $y_3$ and $y_5$
against $W$, and then $\Phi_1$ can be found from eqs.(9) and (15).  The
limiting surface is obtained by setting $U = 0$, where, as before, $U =
\Phi_k + \delta\Phi_0S_0 + \delta\Phi_2S_2 + Q$.

Of greatest interest are those systems in which the edge of the cluster
coincides with the last closed zero-velocity surface.  For a fixed value
of $W_c$ this can be found by iterating the value of $\rho_1/\rho_c$ until a
point on the $x_1$-axis is found at which both $U$ and $\partial U/\partial
x_1$ vanish.  (This point is the Lagrangian point.)

We now discuss the special features of the problem discussed in \S2.2. In the
disk model of the Galaxy, the values of $\alpha_1$ and $\alpha_3$
were calculated at sun's distance using
$A = 0.0144$ km/sec/pc and $B = -0.012$ km/sec/pc with
the density at the solar distance in the Galaxy
$$
\rho_G = 0.11 M_\odot/pc^3,
$$
(cf. Kuijken \& Gilmore 1989).
Using eqs.(25) and (26)
we obtain
$$
{\alpha_3\over\alpha_1} = -3.99,
$$
and so  $S_0 = -4.49$, by eq.(27). Again $S_2$ is largest along the $x_1$-axis
and smallest
along the $z_1$-axis:
the relevant values are $S_2(\pi/2,0) = 3.00$,
$S_2(\pi/2,\pi/2) = +1.50$ and $S_2(0,0) = -4.49$.
Comparison with the values of $S_2$ given
for the point-mass model indicate that the effects will be more
pronounced in the case of the disk model.

\noindent {\bf4. Results and Discussion}

\noindent4.1 Description of the models

Space density profiles of the resulting models are shown in Figs. 1-3 and Figs.
6-8 for the point-mass and the disk model of the galaxy, respectively. The
space
density profiles along the $x_1$-axis are compared with the corresponding King
models in Figs. 1 and 6. Figs. 4, 5, 9 and 10 show the surface densities of our
model clusters along the $x_1$- and $z_1$-axes as seen from the $y_1$
direction.
It can be seen from the results that:

1. The clusters are triaxial, with the longest axis in the
direction to the galactic centre and the shortest in the direction
at right angles to the orbital plane. Denoting by $a,b,c$ the semi
diameters in the $x_1,~y_1$ and $z_1$ directions, respectively, it is found
that the axial ratios $a/b$ and $b/c$ decrease only slightly as the scaled
central potential $W_c$ increases from 2.5 to 10. Typical values
are $a/b=1.5, b/c=1.04$ for the point mass perturber and $a/b=1.50,
b/c=1.34$ for the tidal field representing a disk galaxy. The strong
flattening in the $z_1$ direction in the latter case is already known
from $N$-body simulations (Terlevich 1987). The axial ratio will
clearly depend on the values of $\alpha_3/\alpha_1$.

2. The ratio of the maximum extent of the cluster to that of the
corresponding King model decreases slightly as $W_c$
increases and is almost the same in the case of a point mass
perturber as in that of a disk galaxy: a representative value is 1.5,
i.e. the approximate self-consistent model is 50 percent larger.
Interestingly, a similar conclusion was reached by Spitzer(1987)
for a model in which both the cluster and the galaxy are treated
as point masses. This agreement suggests that departures of the
cluster potential from spherical symmetry are modest and helps
to justify the drastic truncation of the series in equation (15)
at $n=2$.

3. The central regions of the clusters are almost unaffected; they
are nearly spherical and differ little from the central regions of
the corresponding King models.

\noindent 4.2 Discussion

Our result (2 above) that our models are more extended than the
corresponding King models may have significant consequences for the
application of King models to observations of globular clusters. Since
limiting radii have sometimes been inferred by fitting King models, it is clear
that the resulting estimates are likely to be systematically too small.
Furthermore, since the inferred strength of the tidal field or the
inferred mass of the cluster (Freeman 1980, Kontizas \& Kontizas 1983)
depends on the cube of the limiting radius, still larger systematic
errors are liable to be present in the estimates of these quantities.
Nevertheless, it should not be forgotten that the assumptions underlying
both King models and the models discussed in this paper are not strictly
applicable to globular clusters at all, since their galactic orbits are
unlikely to be circular.

{}From the theoretical point of view, our models may be useful for the
construction of initial conditions for $N$-body models of tidally
perturbed clusters.  At present the initial conditions used in such
simulations are spherically symmetric, and over the first few crossing
times the system presumably settles into a quasi-equilibrium in the
non-spherically symmetric tide.

In order to check these ideas a number of $1000$-body simulations have been
conducted using NBODY5 (Aarseth 1985) with three different sets of initial
conditions and parameters: (i) one of the models described in this paper, with
scaled central potential $W_c = 2.5$, and the tidal parameters given at the end
of \S3; (ii) a King model with the same central potential and tidal parameters;
and (iii) the same King model without an external tide.  All models have stars
of equal mass, and units are standard (Heggie \& Mathieu 1986), with the
crossing time being $2\surd2$ units.  Figs.(11)-(13) show, for these three
models, the mean square values of $x_1,~y_1$ and $z_1$, measured relative to
the
density centre (Casertano \& Hut 1985).

These results indicate that the model constructed according to the
theory of the present paper remains in approximate dynamic equilibrium
for about the first 4 time units. (The dip in one of the curves in
Fig.11 is no greater than that in one of the curves in Fig.13,
indicating that it can be ascribed to finite-$N$ effects). After about
$t\simeq4$ there begins a gentle expansion (on a
time scale of about 70 time units). The tidally perturbed King model, on
the other hand, flattens in the $z_1$-direction over the same period of
about 4 units and
spreads in the other two directions.  Indeed, at about $t=4$ it quite
closely resembles the model shown in Fig.11.  Eventually this model
shows the same tendency to expand, and it does so on a very similar time
scale to the model shown in Fig.11.  We interpret the expansion as being
due to two-body relaxation, and this is confirmed by the isolated model,
which shows no tendency to triaxiality, but does expand in the last part
of the run.

In conclusion we remark that it would be desirable to improve the models
by a higher-order expansion, or use of a grid-based potential solver.
Such a method has been used by Rix and White (1989) to discuss models
for a binary galaxy. The models obtained by us can be regarded as a
limiting and simplified case of the same problem.

\bigskip\noindent
{\bf Acknowledgements}
\smallskip

The idea of using a distribution function depending on $J$ arose during
a conversation with Dr S.D.M. White.  The stimulus for finishing this
paper came from Dr S.J. Aarseth, but the research was mainly carried out
under a grant (No. SGC 25013, 1986-88) from the UK Science and
Engineering Research Council, whose support we gratefully acknowledge.
NR also acknowledges the use of computing equipment at the Royal
Greenwich Observatory.

\bigskip\noindent
{\bf References}

\leftskip=0.3truein
\parindent=-0.3truein

Aarseth S.J., 1985, in Brackbill J.U., Cohen B.I., eds, Multiple
Time Scales. Academic Press, New York, p. 377

Casertano S., Hut P., 1985, ApJ, 298, 80

Chandrasekhar S., 1942,  Principles of Stellar Dynamics. University
  of Chicago Press, Chicago

Freeman K.C., 1980, in Hesser J.E., ed, Proc. IAU Symp. 85, Star clusters.
Reidel,
 Dordrecht, p.317

Gunn J.E., Griffin R.F., 1979, AJ, 84, 752

Heggie D.C., Mathieu R.D., 1986, in  Hut P.,  McMillan S., eds,
The Use of Supercomputers in Stellar Dynamics. Springer Verlag,
Berlin, p.233

Heggie D.C., Ramamani N., 1993, in Meylan G., Djorgovski S., eds, Dynamics of
Globular Clusters. PASP Conference Series, San Francisco, in press

Illingworth G.,  Illingworth W., 1976,  ApJS, 30, 227

King I.R., 1966,  AJ,  71, 64

Kontizas E., Kontizas M., 1983,  A\&AS, 52,  143

Kuijken K.,  Gilmore G., 1989, MNRAS, 239, 651

Meylan G., 1988,  A\&A, 191, 215

Rix H-W.R., White S.D.M., 1989, MNRAS, 240, 941

Spitzer L., Jr., 1987,  Dynamical Evolution of Globular Clusters.
 Princeton University Press, Princeton, p.104

Terlevich E., 1987, MNRAS, 224, 193

Weinberg M.D., 1993, in Brodie J.P., Smith G.H., eds,  The
Globular Cluster--Galaxy Connection. PASP Conf. Ser., San Francisco, in press
\vfill\eject
\parindent=0pt
\leftskip=0pt
{\bf Figure Captions}
\medskip
Fig.1 Space density on the $x_1$-axis for nine of the models described
in this paper and the corresponding King models, in the case in which
the tidal field is due to a large distant point mass.  The scaled
central potentials $W_c$ (denoted $W_0$ on the figure) of the models are
$2.5,~3,~4,~5,~6,~7,~8,~9$ and $10$, and the King models are
distinguished by dotted curves.  The scaled distance from the cluster
centre is $R$, and the density is scaled by the central value.

Fig.2 As Fig.1, but for the $y_1$-axis. The King models are not shown.

Fig.3 As Fig.1, but for the $z_1$-axis.

Fig.4 The projected (surface) density of the nine models displayed in
Figs.1-3. The surface density is scaled to the central value. The cluster is
viewed along the $y_1$-axis and the profile along the $x_1$-axis is plotted.

Fig.5 As Fig.4, except that the profile along the $z_1$-axis is plotted.

Fig.6 As Fig.1, except that the tidal field is a model of that in the
solar neighbourhood (``disk'' case).

Fig.7 As Fig.2, but for the disk case.

Fig.8 As Fig.3, but for the disk case.

Fig.9 As Fig.4, but for the disk case.

Fig.10 As Fig.5, but for the disk case.

Fig.11 Evolution of a $1000$-body model with initial conditions obtained
from one of the models described in this paper.  The model had scaled
central potential $W_c = 2.5$ and the tidal field is a model of the tide
in the solar neighbourhood.  The quantities plotted are the mean square
values of the coordinates $x_1,~y_1$ and $z_1$, but using as origin the
``density centre''.  The lowest curve corresponds to $z_1$, the
uppermost to $x_1$.

Fig.12 As Fig.11, except the initial conditions are constructed using a
standard (spherically symmetric) King model.  Again the lowest curve
corresponds to $z_1$, the uppermost to $x_1$.

Fig.13 As Fig.12, but without any tide. The dashed curve gives one third of the
mean square distance from the density centre, $\langle
x_1^2+y_1^2+z_1^2\rangle/3$.

\bye